\documentclass[12pt]{iopart}
\usepackage{epsfig}
\usepackage{citesort}
\usepackage{color}

\newcommand{\gtrsim}{\,\rlap{\lower3.7pt\hbox{$\mathchar\sim$}}
\raise1pt\hbox{$>$}\,}
\newcommand{\lesssim}{\,\rlap{\lower3.7pt\hbox{$\mathchar\sim$}}
\raise1pt\hbox{$<$}\,}

\usepackage{color}

\definecolor{Black}{named}{Black}
\definecolor{Blue}{named}{Blue}
\definecolor{Red}{named}{Red}

\newcommand{\mix}{\chi_0}
\newcommand{\mixe}{\chi}

\newcommand{\muu}{m_{\gamma^{\prime}}}

\newcommand{\OP}{\omega_\mathrm{_P}}

\renewcommand\({\left(}
\renewcommand\){\right)}

\newcommand{\be}{\begin{equation}}
\newcommand{\ee}{\end{equation}}
\def\bea{\begin{eqnarray}}
\def\eea{\end{eqnarray}}

\begin{document}

\hfill  DESY 08-205, MPP-2008-169

\title{Microwave Background Constraints on Mixing of Photons with Hidden Photons}

\author{Alessandro Mirizzi}
\address{ Max-Planck-Institut f\"ur Physik (Werner Heisenberg
Institut) \\ F\"ohringer Ring 6,
80805 M\"unchen, Germany\\}
\author{Javier Redondo}
\address{Deutsches Elektronen Synchrotron \\
 Notkestra\ss e 85, 22607 Hamburg, Germany\\}
\author{G\"unter Sigl}
\address{II. Institut f\"ur theoretische Physik, Universit\"at Hamburg,
Luruper Chaussee 149, 22761 Hamburg, Germany\\}
\date{16 December 2008}

\begin{abstract}
Various extensions of the Standard Model predict the existence
of hidden photons kinetically mixing with the ordinary photon.
This mixing leads to oscillations between photons and hidden photons,
analogous to the observed oscillations between different neutrino flavors.
In this context, we derive new bounds on
the photon-hidden photon mixing parameters using the high precision
cosmic microwave background spectral
data collected by the Far Infrared Absolute Spectrophotometer
instrument on board of the Cosmic Background Explorer.
Requiring the distortions of the CMB induced by the photon-hidden
photon mixing to be smaller than experimental upper limits, this leads
to a bound on the  mixing angle $\mix\lesssim 10^{-7}-10^{-5}$ for
hidden photon masses between $10^{-14}\,$eV and $10^{-7}\,$eV.
This low-mass and low-mixing region of the hidden photon parameter
space was previously unconstrained. 
\\

\end{abstract}
\noindent 

\maketitle

\section{Introduction}

The last decade has witnessed the blooming of modern cosmology,
supported by a consolidated model of particle physics and 
a huge amount of new observational data with unprecedented precision 
(see for example the latest WMAP team results~\cite{Hinshaw:2008kr,Nolta:2008ih}). 
Every observation has been shown to fit with the predictions of the so-called $\Lambda$-CDM 
model, a spatially flat Friedman universe, with $\sim 10^{-8}$ baryons per photon and whose energy 
composition is now dominated by two unknown ingredients: dark matter and dark energy.
At this stage of concordance, observational cosmology 
can be used to test the existence of non-standard particle physics.

One of the most powerful probes used to constrain exotic physics 
is represented by the observations of the cosmic microwave background radiation.
Standard cosmology predicts relic radiation from the big bang which, being 
originally in thermal equilibrium with matter, suffered a soft decoupling 
and, therefore, still features today a perfect black body spectrum.
As it turns out, during most of the history of the universe this perfect blackbody
has been completely unprotected against distortions. 
Therefore, any new particle beyond the standard model which can interact with 
photons can potentially distort the blackbody spectrum, providing us
with a glimpse of the existence of such a particle.

These distortions are expected to be more severe if the new particles have
feeble couplings, so that their interactions with photons do not allow to establish
thermal equilibrium (but of course strong enough to produce some effect), 
and if they have small masses, such that photons can annihilate into them 
and/or scattering processes are not suppressed by a heavy mass scale.
Since the cosmic radiation spectrum freezes out when the temperature
of the universe falls below the $\sim $ eV scale, the spectrum of the cosmic
microwave background (CMB) can, therefore, be 
an excellent probe for weakly interacting sub-eV particles (WISPs).

Among the most elusive WISPs we find low mass hidden photons (HPs), gauge
bosons of a ${\mathcal U(1)}_{\rm hid}$ gauge symmetry having
kinetic mixing with the ordinary photon. Due to the mixing one expects
photon-HP oscillations driven by the mass difference of the two particles.
Clearly, as this mass becomes smaller the oscillation length grows and 
oscillations are harder to detect. In this respect, the CMB provides an 
excellent probe for photon oscillations into low mass HPs, 
since the beam-line is the longest at our disposal, namely the whole universe.
Indeed, hidden photons were the first WISP candidates to be confronted 
with CMB data in the early 80's~\cite{Georgi:1983sy}.

Recently, new intriguing ideas and  experimental  techniques have
been proposed to achieve possible detections of these still elusive
particles~\cite{Jaeckel:2007ch,Gninenko:2008pz,Jaeckel:2008sz}.
Moreover, the astrophysical and cosmological role of HPs is currently
under exploration~\cite{Redondo:2008aa,Jaeckel:2008fi,Zechlin:2008tj,Postma:2008ec}.

The purpose of this paper is to revisit the CMB bounds on photon-HP
mixing using the most precise available observations of the CMB spectrum, 
those provided by the Far Infrared Absolute Spectrophotometer (FIRAS) on board 
of the Cosmic Background Explorer (COBE)~\cite{Fixsen:1996nj,Mather:1998gm}. 
The high precision of this measurement, which confirmed the blackbody 
nature of the spectrum at better than 1 part in 10$^{4}$, has been already exploited 
to constrain hidden photons~\cite{Nordberg:1998wn} and also 
other WISPs such as axions~\cite{Mirizzi:2005ng}, or radiative
neutrino decays~\cite{Mirizzi:2007jd} or millicharged 
particles~\cite{Melchiorri:2007sq}. 
However, the already existing constraints~\cite{Georgi:1983sy,Nordberg:1998wn} 
did not properly take into account the refractive properties of the primordial plasma 
and are thus incomplete. The medium effects are especially important when we realize that 
a resonant photon-HP conversion is possible, similar to the Mikheev-Smirnov-Wolfenstein 
(MSW) effect in the neutrino case~\cite{Wolfenstein:1977ue,Mikheyev:1985aa,Mikheyev:1985bb}.
This resonant conversion is much stronger than the vacuum oscillations considered
so far and, therefore, the bounds we obtain are much stronger.

The plan of our work is as follows.
In Section~2 we summarize the relevant formalism concerning the photon-HP mixing. 
In particular, we describe how the medium effects modify the oscillations 
in our system. In Section~3 we present our analytical recipe to calculate the 
photon-HP conversion probability across a resonance in the expanding universe.
In Section~4 we describe our simplified model for the effective photon mass, 
induced by the primordial plasma.
In Section~5 we describe the constraints coming from spectral CMB distortions. 
In Section~6 we outline in a general way the strength of photon-HP
mixing bounds that can be obtained from an observed modification of the 
photon flux from a generic astrophysical source.
Finally, in Section~7 we  compare our new cosmological bound with the
other ones existing in the literature and we draw our conclusions.

\section{Photon-Hidden Photon Mixing}

A hidden photon is the gauge boson of a gauge ${\mathcal U(1)}_{\rm hid}$ symmetry under
which all the SM fields are uncharged, and thus remains hidden in our world.
At energies above the electroweak scale, particles charged under ${\mathcal U(1)}_{\rm hid}$
and the SM hypercharge are likely to exist and act as messengers between the two
``sectors" of low-energy physics.
Their effects will produce effective operators in the low-energy theory
that contain the SM and the hidden photon.
Typically these operators are irrelevant, suppressed by the heavy particle masses.
But there is still one marginal operator, the so called kinetic 
mixing~\cite{Holdom:1985ag}
whose effects are not in principle suppressed by heavy masses and
thus can lead to relatively large effects.
The low energy Lagrangian of such an extension of the SM would be then~\cite{Jaeckel:2008fi}
%........................................................
\begin{eqnarray}
{\cal L} &=& -\frac{1}{4}F_{\mu \nu}F^{\mu \nu}
 - \frac{1}{4}B_{\mu \nu}B^{\mu \nu} \nonumber \\
&+& \frac{\sin\mix}{2} B_{\mu \nu} F^{\mu \nu}
+ \frac{\cos^2\mix}{2} \muu^2 B_{\mu}B^\mu + j^\mu_{\rm em} A_{\mu} \,\ ,
\end{eqnarray}
%....................................................................
where $A_{\mu},B_\mu$ are, respectively, the photon and the hidden photon fields,
$F_{\mu\nu}$ and $B_{\mu\nu}$ their respective field strengths and $j_{\rm em}^\mu$
is the electromagnetic current. We have included a hidden photon mass
$\muu$ which we will treat as a free parameter.
Typical predicted values for the mixing angle $\mix$ in realistic string compactifications
range between $10^{-16}$ and $10^{-2}$~\cite{Dienes:1996zr} (see also~\cite{Abel:2006qt,Abel:2008ai}).

The kinetic mixing term can be removed, leading to a canonical form of the
kinetic lagrangian, by the following change of basis
%................................................................
\begin{equation}
\{A,B\}\rightarrow\{A_{R},S\} \,\ ,
\end{equation}
where
\begin{eqnarray}
A_{R}&=&\cos\mix A \,\ , \\
S&=& B-\sin\mix A  \,\ .
\end{eqnarray}
%......................................................
 Let us call $\gamma,\gamma_s$ the quanta of the $A_R$ and $S$
fields, respectively.
The basis $\{\gamma,\gamma_s\}$ can be called the ``interaction basis",
since the photon $\gamma$ is the
 state that interacts with SM charged particles and $\gamma_s$, being
orthogonal to $\gamma$, is completely sterile.
In the interaction basis the kinetic lagrangian is diagonal but the
mixing angle $\mix$ appears in an off-diagonal term in the mass-squared matrix,
%.....................................................

\be
\mathcal{M}^2 =
\left(
\begin{array}{cc}
\muu^2 \sin\mix^2 & \muu^2  \sin\mix\cos\mix     \\
\muu^2\sin\mix\cos\mix   & \muu^2 \cos^2\mix
\end{array}
\right) \ \ .
\ee

%................................................
The mass matrix $\mathcal{M}^2$ can be diagonalized through the unitary
matrix $U$
%..............................................
\begin{equation}
U=\left( \begin{array}{c c}
\cos\mix & - \sin\mix \\
\sin\mix & \cos\mix
\end{array}
\right) \,\ ,
\end{equation}
%.............................................
that allows to identify the two ``propagation states"
%..................
\begin{equation}
\left( \begin{array}{c}
\gamma_1 \\
\gamma_2
\end{array}
\right) =
U
\left( \begin{array}{c}
\gamma \\
\gamma_s
\end{array}
\right) \,\ ,
\end{equation}
%.......................................
where $\gamma_1$ is mostly photon-like and massless, while
 $\gamma_2$ has mass
 $\muu$ and is close to the sterile state.
The mismatch between the interaction $\{\gamma,\gamma_s\}$
and propagation $\{\gamma_1,\gamma_2\}$ states by the  mixing angle
$\mix$ is well-known to produce $\gamma\to\gamma_s$ oscillations~\cite{Okun:1982xi},
with a conversion probability in vacuum given by
%..................................................................
\begin{equation}
P_{\gamma\to\gamma_s} = \sin^2 2\mix\sin^2(\muu^2 L/4\omega) \,\ ,
\label{eq:vacuum}
\end{equation}
where $L$ is the path length and $\omega$ is the photon energy.

In analogy to the neutrino case~\cite{Kuo:1989qe}, the $\gamma\leftrightarrow\gamma_s$ oscillations are modified
by the refraction properties of the medium.
In the primordial plasma, photons acquire a non-trivial dispersion relation
 which can be parametrized by adding an effective photon mass $m_\gamma$ to the 
Lagrangian.
 This mass depends on the properties of the medium, as we will show explicitly in
Section~4.
 This is generally complex, reflecting the absorption properties of the plasma,
but in the context relevant for the present paper the imaginary part is negligible.
In this case, the effective mixing angle is related to the vacuum
by~\cite{Jaeckel:2008fi,Redondo:2008aa}
%................................................................
\begin{eqnarray}
\sin 2\mixe &=& \frac{\sin 2\mix}{\sqrt{\sin^22\mix+\(\cos2\mix-\xi\)^2}} \,\ , \nonumber \\
\cos 2\mixe &=& \frac{\cos2\mix-\xi}{\sqrt{\sin^22\mix+\(\cos2\mix-\xi\)^2}} \,\ ,
\label{eq:mixmatt}
\end{eqnarray}
%..............................................................
where the parameter
%...............................
\begin{equation}
\xi= m^2_\gamma/\muu^2
\end{equation}
measures the significance of the medium effects.

%%%%%%%%%%%%%%%%%%%%%%%%%%%%%%%%%%%%%%%%%%%%%%%%%%%%%%%%%
\section{Photon-Hidden Photon Oscillations in the Expanding Universe}

%%%%%%%%%%%%%%%%%%%%%%%%%%%%%%%%%%%%%%%%%%%%%%%%%%%%%%%

To obtain the bound on photon-hidden photon mixing, we have
 to compute the fraction of CMB photons that would oscillate into invisible $\gamma_s$.
The photon effective mass squared $m^2_\gamma$ is generally proportional to the density
of charged particles in the medium, so it relaxes as the photons propagate in the expanding
primordial plasma.
We are then facing a complicated problem of oscillations in an inhomogeneous medium.

For sufficiently early times, $m_\gamma \gg \muu$ and, 
therefore, the photons are very close to be
both interaction and propagation states at a given time. In this case, 
oscillations into $\gamma_s$ are suppressed, 
%........................................
\begin{equation}
\xi\gg 1 \,\ , \,\ \,\  \,\ \,\ \,\ {} \mixe\to \pi/2 \,\ \,\ \,\ \textrm{(medium suppression)} \,\ .
\end{equation}
%......................................
As the universe expands, $\xi$ eventually reaches
%...................................................
\begin{equation}
\xi = \cos\mix \,\ , \,\ \,\ \mixe\to \pi/4 \,\ \,\ \,\ \textrm{(resonance condition)} \,\ .
\end{equation}
When this condition is fulfilled,
 resonant photon-hidden photon conversions are possible, analogous to the well-known
Mikheev-Smirnov-Wolfenstein (MSW) effect in the neutrino case~\cite{Wolfenstein:1977ue,Mikheyev:1985aa,Mikheyev:1985bb}.
As $\xi$ decreases below unity,
\begin{equation}
\xi \ll  1 \,\ , \,\ \,\ \,\ \,\ \,\  \mixe\to \mix \,\ \,\ \,\ \,\ \,\ \textrm{(vacuum oscillations)} \,\
\end{equation}
 $\gamma\to\gamma_s$ oscillations  take place as if
they occurred in vacuum.

In vacuum, oscillations are proportional to $\sin^22\mix$ [see Eq.~(\ref{eq:vacuum})] and since the blackbody nature of the
CMB is established experimentally to an accuracy of $\simeq 10^{-4}$ we can easily
 exclude mixing angles $\gtrsim 10^{-2}$.
However, during a resonance $\sin 2\mixe \sim 1$ and
the conversion $\gamma\to \gamma_s$ can be much stronger.
In this paper we want to focus on the range of
 HP masses that can undergo such a resonant transition, and are,
therefore, strongly constrained.

When the photon production and detection points are separated by many
oscillation lengths, then on both sides of a resonance,
$m_\gamma=\muu$, the oscillation patterns wash out and the
transition probability is given by~\cite{Parke:1986jy}
\be
P_{\gamma\to\gamma_s}=\frac{1}{2}+\(p-\frac{1}{2}\)\cos 2\mix\cos2\mixe \,\ ,
\label{P}
\ee
where $\mix,\mixe$ are the mixing angles at the detection and production points
considered to be
in vacuum and high density, respectively, and $p$ is the level crossing probability.
This latter takes into account the deviation from adiabaticity of photon-HP oscillations
in the resonance region.
In particular, one has $p=0$ for a completely adiabatic transition
and $p=1$ for an extremely nonadiabatic one.

As we will see, we are going to bound mixing angles much smaller than $\mix \sim 10^{-2}$, so
for simplicity we can already take $\cos 2\mix\simeq -\cos 2\mixe =1$ 
in Eq.~(\ref{P}) which,
therefore, takes the much simpler form
\be
P_{\gamma\to\gamma_s}\simeq 1-p \ .
\label{PP}
\ee
The FIRAS sensitivity will allow us to bound $1-p\lesssim 10^{-4}$,
requiring thus a non-adiabatic resonance.

The crossing probability $p$ for photon-hidden photon resonant conversions
can be obtained using the  Landau-Zener expression
%.........................................
\begin{equation}
p \simeq \exp(-2\pi r k \sin^2 \mix) \,\ ,
\label{eq:landau}
\end{equation}
%.......................................
where $k=\muu^2/2\omega$ is the $\gamma\to\gamma_s$ vacuum oscillation
wavenumber and
%..............................................
\begin{equation}
r=\left|\frac{d \ln m^2_\gamma(t)}{dt}\right|^{-1}_{t=t_{\rm res}}
\label{eq:r}
\end{equation}
%........................................................
is a scale parameter to be evaluated at the location where
a resonance occurs.

This expression has been widely used in solar and supernova
neutrino oscillations~\cite{Kuo:1989qe}.
It represents an accurate ansatz to calculate the level crossing
probability for  general density profiles (see, e.g. \cite{Fogli:2003dw}).
In particular, we observe that it reproduces the correct limits $p\simeq0$
for $r\to \infty$ (adiabatic limit)
and $p\simeq1$ for $r\to 0$ (extreme non-adiabatic limit).

Finally, under the approximations outlined before, we can approximate
the non-adiabatic conversion probability as
\be
P_{\gamma\to\gamma_s}\simeq 2\pi r k \mix^2 = \frac{\pi \muu^2 \mix^2 }{\omega} \left|
\frac{d \ln m^2_\gamma(t)}{dt}\right|^{-1}_{t=t_{\rm res}} \ .
\label{PPP}
\ee
The half-width width of the resonance is, according to Eq.~(\ref{eq:mixmatt}), 
$\delta \xi(t)\simeq \sin 2\mix$, which corresponds to a time scale
\be
\tau_r \simeq  r \sin2\mix \,\ .
\label{taur}
\ee
Due to the smallness of the vacuum mixing angle, the
resonance is very narrow, so that it is reasonable to take into account the
deviation from the adiabaticity only at the crossing point.
%.............................................................
\section{Cosmological $m_\gamma$ profile}
%............................................................

From Eq.~(\ref{PPP})
it is clear that the only information needed to calculate the conversion probability
is the profile of the photon effective mass $m_\gamma$
along the cosmological line of sight.
The primordial plasma is thought to be electrically neutral and
composed mainly of hydrogen and helium in a fraction per mass
$Y_p=m_{\rm He}/m_{\rm H}\simeq 0.25$.
The effective mass squared has a positive and a negative contribution~\cite{Born:1980}
from scattering off free electrons and off neutral atoms
%
%%%%%%%%%%%%%%%%%%%%%%%%%%%%%%%%%%%%%%%%%%%%%%%%%
\footnote{The vacuum magnetic birefringence~\cite{Adler:1971wn} due to a 
primordial magnetic field
is negligible~\cite{Yanagida:1987nf}.}
%%%%%%%%%%%%%%%%%%%%%%%%%%%%%%%%%%%%%%%%%%%%%%%%%
%
%
\bea
m_\gamma^2 &\simeq&
 \OP^2-2\omega^2(n-1)_{\rm H} \nonumber\\
&\simeq& 1.4\times 10^{-21}
\left( X_e- 7.3 \times 10^{-3} \left(\frac{\omega}{\mathrm{eV}}\right)^2(1-X_e)
\right)\frac{n_p}{\mathrm{cm^{-3}}}\ \mathrm{eV}^2 \ ,
\label{pm}
\eea
where $\OP^2= 4 \pi \alpha n_e/m_e$ is the 
plasma frequency with $\alpha$ the fine structure constant, 
$m_e$ the electron mass and $n_e$ the free electron density.
 We have written the ionized fraction of hydrogen as
 $X_e=n_e/n_{p}$ with $n_p$ the proton density.
The indices of refraction of neutral hydrogen and
helium are $(n-1)_{\rm H}=13.6\times 10^{-5}$
 and $(n-1)_{\rm He}=3.48\times 10^{-5}$ in normal conditions~\cite{Born:1980},
 rather insensitive to $\omega$. Since the fraction of electrons
corresponding to helium is $\sim13 \%$ and $(n-1)_{\rm He}\ll(n-1)_{\rm H}$ we have neglected the effects of helium in
 Eq.~(\ref{pm}).

The photon frequency $\omega$ and the proton density $n_p$ are
given in terms of redshift $z$, the photon energy today $\omega_0$, 
the CMB temperature
today $T_0$ and the baryon to photon ratio 
$\eta\simeq 6.7 \times 10^{-10}$ as~\cite{Kolb:1990vq}
\be
\omega=\omega_0(1+z) \,\,\, ; \,\,\, n_p = \left(1-\frac{Y_p}{2}\right)\eta \frac{2\zeta(3)}{\pi^2}T_0^3(1+z)^3 \,\ .
\ee
Note that for a negative effective mass squared a resonance is \emph{not} possible.
However, the negative contribution is proportional to $\omega^2(1-X_e)$ and thus
is unimportant for sufficiently small frequencies  (small $\omega_0$ or late times
when the redshift $z$ is small) and/or large ionization fractions, $X_e\simeq 1$.
Interestingly, the smallest frequencies are the most precisely determined by FIRAS.

Since the conversion probability Eq.~(\ref{PPP}) is proportional to $r$ and $\omega^{-1}$
which both grow with decreasing redshift, \emph{for a fixed HP mass} the 
later the resonance the more adiabatic the transition will be and the stronger our constraints
on the mixing parameter $\mix$ will be.

\begin{figure}[t]
\begin{center}
\includegraphics[width=6.4cm]{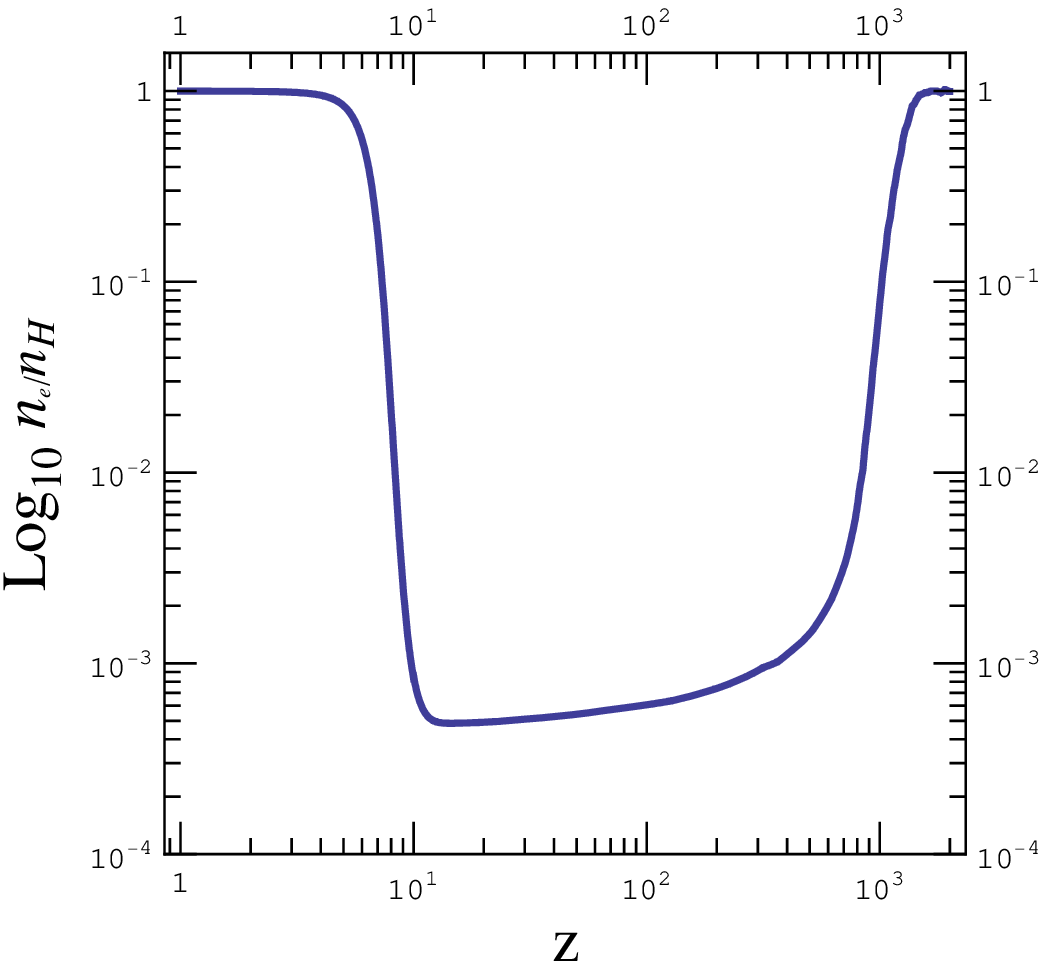}\hspace{.5cm}\includegraphics[width=8cm]{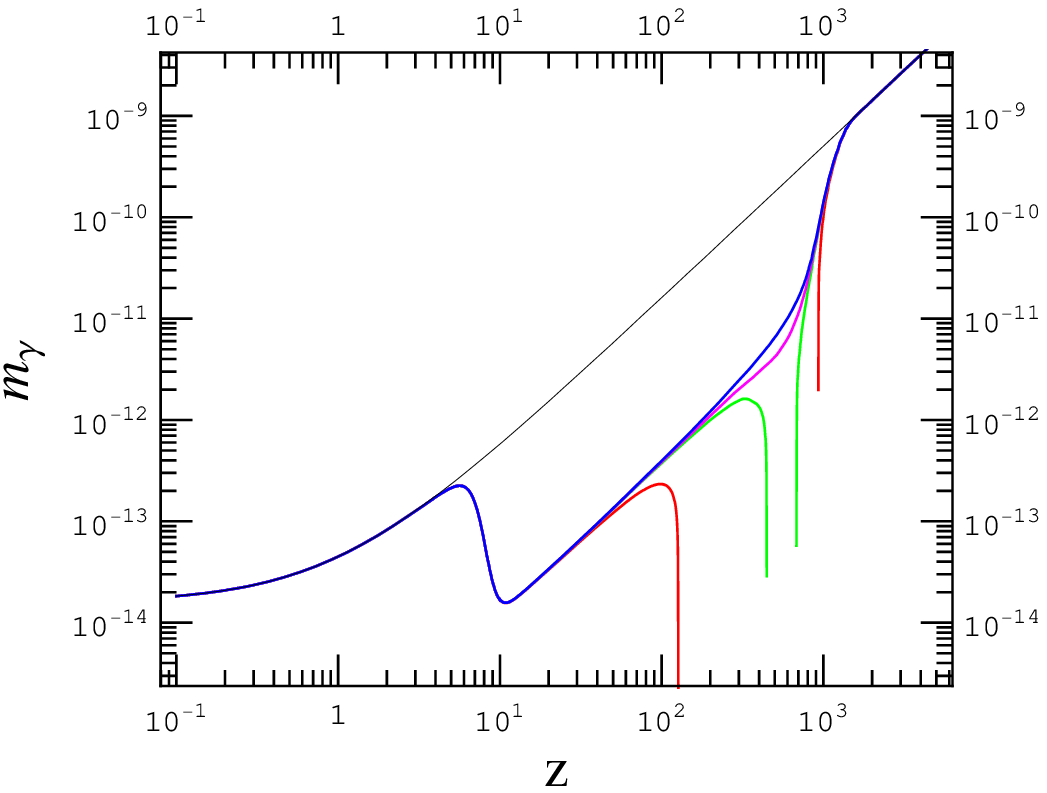}
\caption{Left panel: The ionization fraction of hydrogen. For redshifts above $z=11$ the line is taken from the data of~\cite{Seager:1999bc}. The re-ionization period between $z=6$ and $z=11$ is modeled.
Right panel: Effective photon mass as function of redshift. The thin line
 uses $X_e=1$, the blue, magenta, green and red lines are for $\omega/T=1,3,4,10$,
 respectively. 
The two sharp dips bound the region where $m_\gamma$ becomes imaginary.}
\label{Xe}
\end{center}
\end{figure}

The history of the ionization fraction $X_e(z)$ is extremely complex.
Above a temperature $T\sim 0.5$ eV (redshift $z\sim 1100$)
hydrogen is fully ionized. As the universe temperature decreases,
  photons cannot ionize hydrogen efficiently and electrons and protons slowly combine.
This makes the universe very transparent to radiation, indeed releasing
the photon bath which we see today as the CMB.
This epoch of so-called recombination has been studied in great detail
 in~\cite{Seager:1999bc} from which we can take the values of $X_e$ as a function
of redshift, shown in Fig.~\ref{Xe}.
Later on, the universe becomes ionized
again due to ultraviolet radiation from the first quasars or population III stars.
 The 5 year data of the WMAP mission
 constrains the redshift of an instantaneous reionization
to $z=11.0\pm1.4$ with $68\%$ confidence level by CMB polarization
 studies~\cite{Hinshaw:2008kr,Nolta:2008ih}.
At the same time reionization can be studied
with the spectra of high redshift quasars, which show
 so-called Gunn-Peterson troughs due to absorption of light at Lyman-$\alpha$ frequencies.
The study of several quasars by the Sloan Digital Sky Survey showed that
 reionization should end around a redshift $z\sim 6$,
 and, therefore, should be an extended process taking place between
$z\sim 6-11$~\cite{Dunkley:2008ie}.

%%%%%%%%%%%%%%%%%%%%%%%%%%%%%%%%%%
%%%%%%%%%%%%%%%%%%%%%%%%%%%%%%%%%%

In Fig.~\ref{Xe} we show a possible profile for the cosmological history
of $X_e$ and $m_\gamma$ as a function of redshift z.
We can learn several important things from this figure.
Let us focus on the right panel. 
We can first compare the thin line, which corresponds to a cosmology when 
atoms are always ionized, i.e. $X_e=1$, with the more realistic 
colored 
lines which includes the ionization history of the left panel and assumes
 $\omega/T=1,3,4,10$, respectively.
We see that during the dark ages ($6\lesssim z\lesssim1000$), 
$X_e$ drops very much and the resonances  are moved to higher redshifts 
compared with the $X_e=1$ case. 
As we commented before, the later a resonance the stronger it is,
and, therefore, the inclusion of the ionization history tends to decrease 
the bounds on $\mix$.

On the one hand, due to the smallness of $X_e$ during the dark ages,
the index of refraction of neutral hydrogen can dominate over the contribution
from free electrons turning the effective photon mass squared 
negative (see Eq.~(\ref{pm})).
Since this negative contribution is proportional to the photon frequency squared,
the region of redshifts for which this happens is broader for higher frequencies.
From Fig.~\ref{Xe} we see that this does not happen for $\omega/T\lesssim 3$ but
for $\omega/T=4$ it forbids a resonance in the range of redshifts
 $400\lesssim z \lesssim 700$ 
and for $\omega/T=10$ the range without resonance is enlarged to 
$100\lesssim z \lesssim 1000$. 

Let us recall that this \emph{does not forbid a resonant transition}
but makes it happen at higher redshifts than those which correspond to 
$\omega/T\simeq 1$.
Consider as an example the case $\muu=4\times 10^{-12}$ eV. 
For frequencies $\omega/T\lesssim 2$ the resonance,
$m_\gamma=\muu$, occurs at $z\sim 400$, for $\omega/T = 4$ at
$z\sim 700$ and for $\omega/T= 10$ at $z\sim 1000$.
Moreover, for these delayed resonances it becomes clear that 
the value of $r$, which is related to the inverse of the derivative 
of $m_\gamma$ is smaller than for resonances for low $\omega/T$.
Because of these two facts, the resonances for high $\omega/T$
photons during the dark ages will be much less adiabatic 
than those for lower $\omega/T$ and, therefore, they will play a minor 
role in our bounds.
There is only one possible exception to this conclusion. For very small masses 
$\muu\lesssim 10^{-14}$ eV there is no resonance at all for low $\omega/T$
but still there are resonances for $\omega/T\gtrsim 4$ during the dark ages. 
We have checked however that such resonances are extremely non-adiabatic and no 
interesting bounds can be derived from them. 
Therefore, in the following discussion we will not discuss
resonances for $\omega/T \gtrsim 4$, although we will include them 
in our bounds.

Regarding the resonant conversion $\gamma\to\gamma_s$, 
and focusing on $\omega/T\lesssim 3$, 
we see three different possibilities in the homogeneous universe:
\begin{itemize}
\item $\muu \gtrsim 2\times 10^{-13}$~eV: there is only one resonance crossing
\item $2\times 10^{-13}$~eV $\lesssim  \muu \lesssim 10^{-14}$~eV:
 there are three level crossings respectively after, during and
before reionization. 
\item $\muu\lesssim 10^{-14}$ eV: no resonance is possible.
\end{itemize}

Our analytical ansatz for the crossing probability [Eq.~(\ref{eq:landau})]
 has been shown to reproduce accurately the transitions also
for nonmonotonic density profiles, where multiple crossings could arise
(see, e.g., the supernova neutrino case~\cite{Fogli:2003dw}).
In this case, assuming the factorization of $n$ different crossings,
one obtains~\cite{Fogli:2003dw}
%........................................
\begin{equation}
p = \frac{1}{2}\left[1- \prod_{i=1,\cdots,n} (1-2 p_i) \right] \,\ ,
\label{eq:multiple}
\end{equation}
%......................................
and, therefore, assuming $1-p\lesssim 10^{-4}$ we can write
%........................................
\begin{equation}
P_{\gamma\to\gamma_s}\simeq 1-p \simeq \sum_i P^i_{\gamma\to\gamma_s} \,\ ,
\label{eq:sum}
\end{equation}
%......................................
where the conversion probability for a resonance is still given by Eq.~(\ref{PPP}).
As expected, since we know that photon depletion has to be very small,
(and, therefore, also a possible re-creation) the dissappearence probabilities at
different crossings just add up.

As commented previously, 
in the presence of multiple crossings ($2\times 10^{-13}$~eV $\lesssim  \muu \lesssim 10^{-14}$~eV),
the relevant crossing is the one occuring after reionization (at $z<6$), since the other
two will be less adiabatic. This situation further simplifies our calculations. 
Since the crossings during reionization do not play a role, 
we need to evaluate them only for $z\lesssim6$ (for $\muu < 2\times 10^{-13}$ eV)
and for, say, $z\gtrsim70$ (for $\muu > 2\times 10^{-13}$ eV).

During the epoch of matter domination,
the small density inhomogeneities already present
at CMB decoupling grow and the density profile along a
line of sight can be quite complicated at the smallest redshifts,
with lots of resonances at galactic length scales.
However, according to Eq.~(\ref{eq:sum}) further resonances tend
to increase the transition probability,
thus tending to strengthen the constraints. It is, therefore, conservative to restrict
oneself to the most adiabatic resonances in the smooth background, represented
in Fig.~1. 

Furthermore, the optical depth $\tau$ for crossing an object of typical size 
$l$ with number density $n$ along a cosmological line of sight of length $s\sim4\,$Gpc is
$\tau\sim nl^2s\simeq0.004(10^{-2}/{\rm Mpc}^{-3})(l/10\,{\rm kpc})^2$ which is
small for galactic objects.
Within an order of magnitude this is consistent with galaxy number counts which yield
$\sim10^8\,{\rm sr}^{-1}$~\cite{Metcalfe:1995}:
Given that one galaxy subtends a solid angle
$\Omega\sim(10\,{\rm kpc}/10^3\,{\rm Mpc})^2\,{\rm sr}\sim10^{-10}\,{\rm sr}$, this
 gives $\tau\sim0.01$.
Similarly, counts of galaxy cluster with mass $\gtrsim10^{15}\,M_\odot$ yield
$\sim10^{-6}\,{\rm Mpc}^{-3}$~\cite{Jenkins:2000bv},
or $\sim10^4\,{\rm sr}^{-1}$. With a typical solid angle subtended
of $\Omega\sim(1\,{\rm Mpc}/10^3\,{\rm Mpc})^2\,{\rm sr}\sim10^{-6}\,{\rm sr}$,
this also gives
$\tau\sim0.01$ for the average number of galaxy
clusters crossed by a given line of sight.

%................................................................
\subsection{Adiabaticity parameter}
%...............................................................

Given our density profile it is now straightforward to
compute the adiabaticity parameter $r$ for each crossing
and from it, the photon disappearance probability.
For masses $\muu < 10^{-14}$ eV this would be very much
 dependent on the inhomogeneities at small redshifts, and we prefer
 not to treat this case. We are left with HP masses above the average 
effective cosmological plasma mass of photons at zero redshift.

Recall that in general $r\propto t$ so the most adiabatic crossing 
is the latest, when the universe expands slower and, therefore, sweeps 
smaller ranges of $m_\gamma$ in a given time. This argument also 
allows us to discard the effect of spatial inhomogeneities unless they have
sizes comparable with the size of the universe. It also makes us
 neglect a possible back-reaction during the reionization crossing.

Let us then write,

\be
\frac{d\log m_\gamma^2}{dt}= \frac{d\log m_\gamma^2}{d z} \frac{d z}{dt} \,\ ,
\ee
with
\begin{equation}
\frac{dz}{dt}=-H_0(1+z)\sqrt{\Omega_\lambda+\Omega_{ m}(1+z)^3+\Omega_{ r}
(1+z)^4} \,\ ,
\end{equation}
where we fix $H_0 = 70$~km~s$^{-1}$~Mpc$^{-1}$,
 $\Omega_m = 0.27 $, ${\Omega}_{\lambda} =0.73$,
$\Omega_r=6 \times 10^{-4}$
 consistent with the recent determinations of cosmological
 parameters~\cite{Dunkley:2008ie}.

The function $d X_e/d  z$ can be computed numerically.
 However, since the crossing during reionization does not play a role,
 we need this function only for $z\lesssim6$
(for $\muu < 2\times 10^{-13}$ eV)
 and for $z\gtrsim70$ (for $\muu > 2\times 10^{-13}$ eV).
 Below $z=6$ we can take it to zero, and for $z\gtrsim70$ we can obtain it from the
 following fitting function
\bea
\log_{_{10}} X_e = -3.15\frac{1}{e^{\frac{z-907}{160}}+1} \,\ ,
\eea
valid at the $\%$ level.

%................................................................
\section{FIRAS Bounds}
%.........................................................

The CMB spectrum measured by FIRAS fits extremely well to a black body
spectrum at a temperature $T_0=2.725\pm 0.002$~\cite{Mather:1998gm}.
The energy range of the  CMB spectrum measured by FIRAS~\cite{Fixsen:1996nj}
is $2.84\times10^{-4}\,{\rm eV}\leq \omega_0 \leq 2.65\times 10^{-3}\,$eV, 
corresponding to $1.2\lesssim \omega_0/T_0\lesssim 11.3$.
In that region, the CMB blackbody becomes unprotected to distortions 
below a cosmic temperature $\sim$ keV, which corresponds to a 
photon mass of $\sim 10^{-4}$ eV. On the other hand, today 
the average plasma mass for photons is as low as $2\times 10^{-14}$ eV.
If HPs exist with a mass between these two values they
will be produced resonantly and leave their imprint on the CMB.

The accuracy of  FIRAS constraints $P_{\gamma\to\gamma_s}\lesssim f\simeq 10^{-4}$,
which, using Eq.~(\ref{PPP}) leads to the bound
\be\label{constr2}
  \mix\lesssim\left(\frac{f\omega}{\pi r m_{\gamma^\prime}^2}\right)^{1/2}
  \simeq1.1\times10^{-9}\,f^{1/2}\,\left(\frac{\mu{\rm eV}}{m_{\gamma^\prime}}\right)\,
  \left(\frac{\omega}{{\rm GHz}}\right)^{1/2}\,\left(\frac{\rm pc}{r}\right)^{1/2}\,.
\ee

In order to sharpen this bound, we have considered the distortion of the overall blackbody
spectrum.

To this end we use the COBE-FIRAS data for the experimentally measured
spectrum, corrected for foregrounds~\cite{Fixsen:1996nj}.  Note that
the new calibration of FIRAS~\cite{Mather:1998gm} is within the old
errors and would not change any of our conclusions.  The $N = 43$ data
points $\Phi^{\rm exp}_i$ at different frequencies $\omega_{i}$ are
obtained by summing the best-fit blackbody spectrum  to the
residuals reported in Ref.~\cite{Fixsen:1996nj}.  The errors
$\sigma^{\rm exp}_i$ are also available.  In the presence of
photon-hidden photon conversion, the original intensity of the ``theoretical
blackbody'' at temperature $T$
\begin{equation}
\label{planck}
\Phi^0({\omega},T) = \frac{\omega^3}{ 2 \pi^2}
\big[ \exp (\omega/T )-1 \big]^{-1}
\end{equation}
would be deformed to
$\Phi({\omega},T,\mix,\muu)=\Phi^0({\omega},T)[1-P_{\gamma\to\gamma_s}
({\omega},\mix,\muu)]$.
We then build the reduced chi-squared function
\begin{equation}
\chi_\nu^2(T,\lambda)=\frac{1}{{N}-1}
\sum_{i}^{N}
\bigg[\frac{\Phi^{\rm exp}_i-\Phi({\omega}_i,T,\mix,\muu)}
{\sigma^{\rm exp}_i} \bigg]^2\,.
\end{equation}
We minimize this function with respect to $T$ for each point in the
parameter space  $\lambda=(\muu,\mix)$, i.e.\ $T$ is an
empirical parameter determined by the $\chi_\nu^2$ minimization for each
$\lambda$ rather than being fixed at the standard value
$T_0=2.725\pm0.002$~K.

\begin{figure}[t]
\begin{center}
\includegraphics[width=10cm]{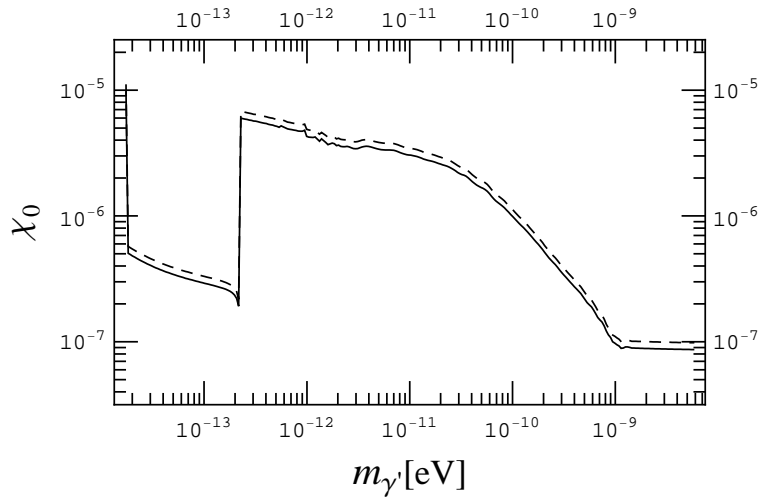}
\caption{Bounds from distortions of the CMB blackbody due to 
$\gamma\to\gamma_s$ photon depletion: 95\% C.L. (solid) and 99\% C.L. (dashed).}
\label{9599}
\end{center}
\end{figure}

In Fig.~\ref{9599} we show our exclusion contour in the plane of $\muu$
and $\mix$. The region above the continuous curve is the
excluded region at 95\% C.L., i.e.\ in this region the chance
probability to experimentally obtain larger values of $\chi_\nu^2$ is lower than~5\%. We
also show the corresponding 99\% C.L. contour which is very close to
the 95\% contour so that another regression method and/or exclusion
criterion would not change the results very much.
Note that for masses $\muu$ for
which the resonant conversion takes place during the dark ages such a resonance is
only possible for $\omega/T\lesssim 2$ for which the positive
contribution to $m_\gamma^2$ dominates over the negative one.
Only the lowest four points in the FIRAS data set satisfy $\omega/T<2$
and are thus distorted.
On the other hand, for larger ($\muu>10^{-10}$ eV, cf. Fig.~\ref{Xe})
 or smaller masses ($2\times 10^{-14}$ eV $<\muu<2\times 10^{-13}$ eV),
 the resonance can happen up to $\omega/T=10$ and we can use the whole FIRAS data set.
 This explains why the bound we obtain is stronger in this latter region.

 Finally, we comment that in our analysis we have assumed that hidden photons are
 produced only by oscillations. In this sense, we have neglected a primordial population
 of hidden photons, that could have subsequently been converted into photons, producing
 an additional distortion of the CMB. Our assumption is reasonable if we neglect the presence
 of primordial charged hidden particles, by whose annihilations hidden photons could have
 been produced.

%.................................................................
\section{Astrophysical bounds}
%.................................................................

The constraint of Eq.~(\ref{constr2}) holds for general astrophysical sources 
whose photon flux at frequency $\omega$ is known to deviate
less than a fraction $f$ from a model prediction in the
absence of mixing with hidden photons.

\begin{figure}[t]
\begin{center}
\includegraphics[width=14cm]{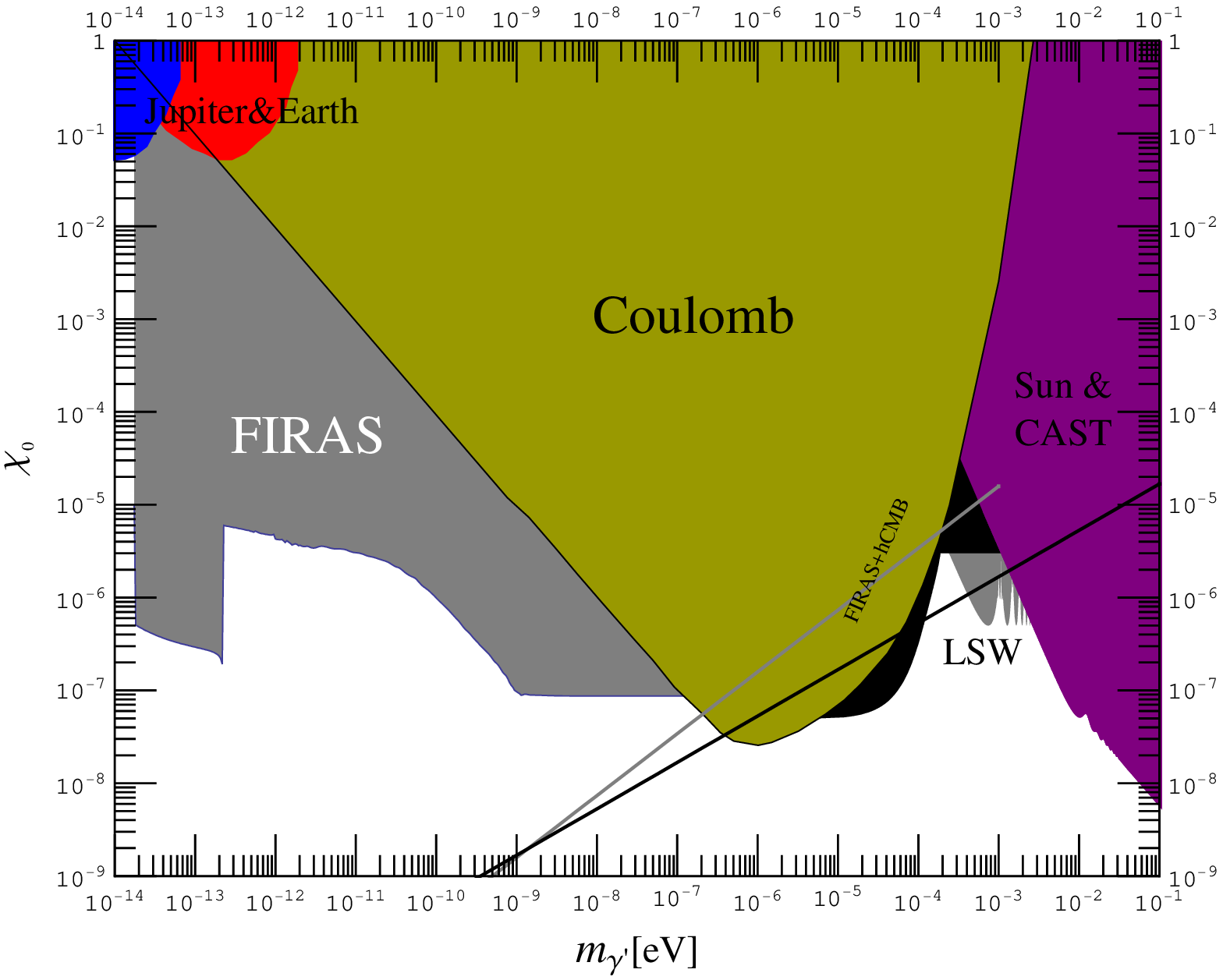}
\caption{Bounds from resonant $\gamma\to\gamma_s$ depletion of the CMB blackbody as constrained by FIRAS data in this work (gray region) and in~\cite{Jaeckel:2008fi} (black region). The gray diagonal line separates the region where the resonance happens at small damping (left, this work) or at strong damping (right, cf.~\cite{Jaeckel:2008fi}).
Plotted for comparison are bounds from tests of the Coulomb $1/r^2$ law~\cite{Bartlett:1988yy,Williams:1971ms}, magnetic fields of Jupiter and earth~\cite{Goldhaber:1971mr},
 photon-regeneration-experiments~\cite{Afanasev:2008fv,Ahlers:2007qf,Ahlers:2007rd,Cameron:1993mr,Chou:2007zzc,Fouche:2008jk}, arguments of the lifetime of the Sun and the CAST search of solar axions~\cite{Redondo:2008aa,Andriamonje:2007ew}.
 The solid black line indicates the
best possible bound Eq.~(\ref{eq:chi}) that can be obtained from 
astrophysical or cosmological
sources whose photon flux is known to be unmodified by photon-HP 
mixing to order unity.}
\label{bounds}
\end{center}
\end{figure}

As long as the contribution of neutrals to $m_\gamma$ is negligible,
the resonance condition is $m_\gamma^2=\OP^2$. 
The resonance thus occurs at the plasma density 
\begin{equation}
\label{eq:resonance}
  n_e\simeq7.3\times10^8\,\left(\frac{m_{\gamma^\prime}}{\mu{\rm eV}}\right)^2\,{\rm cm}^{-3}\,.
\end{equation}
Such resonances can directly influence the photon flux only if the optical depth at the
resonance is smaller than unity. Since the optical depth $\tau_\gamma\gtrsim\sigma_{\rm T}n_e\,r$
with $\sigma_{\rm T}\simeq6.7\times10^{-25}\,{\rm cm}^2$ the Thomson cross section, from
Eq.~(\ref{eq:resonance}) we obtain the condition
\begin{equation}\label{constr3}
  r\lesssim6.6\times10^{-4}\,\left(\frac{m_{\gamma^\prime}}{\mu{\rm eV}}\right)^{-2}\,{\rm pc}
\end{equation}
for the scale over which the mixing potential varies.
Since furthermore only photons with frequency $\omega\gtrsim\OP$ can propagate, we conclude
from Eq.~(\ref{constr2}) that independent of the details of the astrophysical system, only
mixing parameters satisfying
\begin{equation}
\label{eq:chi}
  \mix\gtrsim 5.3 \times10^{-8}  \,    f^{1/2}  \left(\frac{m_{\gamma^\prime}}{ \mu{\rm eV} }  \right)^{1/2}
\end{equation}
can be constrained by observing photon fluxes from astrophysical or cosmological objects.
Note for example that for $m_{\gamma^\prime}\sim$ meV, the best possible bound from direct observations
is thus $\mix\lesssim1.7\times10^{-6}\,f^{1/2}$. Any mixing parameters smaller than this, therefore, has
to be constrained by indirect methods (e.g. stellar cooling) or experimentally.
This theoretical best possible bound from astrophysical
 sources is plotted for $f=1$ in Fig.~\ref{bounds}.
Our CMB based bounds are weaker than this best possible bound essentially because the scale
$r$ in Eq.~(\ref{constr2}) over which the cosmological plasma mass varies is considerably smaller
than the photon mean free path Eq.~(\ref{constr3}) that would lead to the most adiabatic resonance
possible for unabsorbed photons.

%%%%%%%%%%%%%%%%%%%%%%%%%%%%%%%%%%

%%%%%%%%%%%%%%%%%%%%%%%%%%%%%%%%%%

\section{Conclusions}

In this paper, we have revisited the bounds on photon-hidden photon oscillations 
coming from cosmology, deriving updated constraints from
the high precision CMB spectrum data collected by the FIRAS
instrument on board of COBE. We have also commented  about
complementary bounds from a depletion of the photon flux 
from astrophysical sources.
Previous studies~\cite{Georgi:1983sy,Nordberg:1998wn} 
were derived in the pre-COBE  era and/or more importantly 
they lacked a detailed treatment of the effects of the plasma medium
on the photon-hidden photon oscillations.
This has motivated us to re-evaluate the bounds.
This problem has presented interesting analogies to
similar situations encountered in neutrino oscillation
physics. In this regard, we had the benefit to apply
to our case different analytical recipes developed
in neutrino oscillation studies.

The result of our analysis leads
to a bound on the  mixing angle $\mix\lesssim 10^{-7}-10^{-5}$ for
hidden photon masses between $10^{-14}\,$eV and $10^{-7}\,$eV.
In Fig.~\ref{bounds} we plot our FIRAS bound together with the other ones existing in the
literature (see~\cite{Redondo:2008zf} for a complete review). 
It turns out that our new cosmological bound excludes
a region of low-mass and low-mixing angle in the hidden photon parameter space, that
was unconstrained by previous arguments.
As a result of our new bound, it is unlikely that hidden photons with masses
smaller than $10^{-7}\,$eV can play a cosmological role.
Conversely, for meV masses,
resonant photon-hidden photon oscillations happen after nucleosynthesis
 but before CMB decoupling, increasing the effective
 number of neutrinos but also the baryon to photon ratio with interesting
cosmological consequences~\cite{Jaeckel:2008fi}.
The mixing angles  required for this effect could be probed in current laboratory
experiments.

 We  comment that our cosmological bound on hidden photons
 is based on an approximate treatment of the plasma environment.
 This would leave open possible improvements of our limit
when a more accurate description of the primordial plasma will be achieved.
Moreover, further improvements of our  cosmological bound  for extremely 
small masses and mixing
could be reached observing a possible depletion of photons from astrophysical sources,
as we proposed in Sec.~6.
In this context, for example it still remains to investigate if photon-hidden photon oscillations
could have some impact on the apparent dimming of supernovae Ia, mimicking the acceleration
of the universe.

\section{Acknowledgements}
This work was supported by the Deutsche Forschungsgemeinschaft (SFB 676 ``Particles, Strings
and the Early Universe: The Structure of Matter and Space-Time) and by the European Union
(contracts No. RII3-CT-2004-506222).
The work of A.M. is supported by the Italian Istituto Nazionale di Fisica
Nucleare (INFN).
J.R. wishes to thank A.~Ringwald for conversations and the
Max Planck Institute for Physics
in Munich for hospitality during the realization of this work.
We also thank Georg Raffelt for careful reading the manuscript and
for useful remarks on it. 

\clearpage

%%%%%%%%%%%%%%%%%%%%%%%%%%%%%%%%%%%%%%%%%%%%%%%%%%%%%%%%%%%%%%%%%%%%%%

\section*{References} %%%%%%%%%%%%%%%%%%%%%%%%%%%%%%%%%%%%%%%%%%%%%%%%

%%%%%%%%%%%%%%%%%%%%%%%%%%%%%%%%%%%%%%%%%%%%%%%%%%%%%%%%%%%%%%%%%%%%%%

\bibliographystyle{/Users/coliflor/utcaps}
\bibliography{/Users/coliflor/DESY*}

\providecommand{\href}[2]{#2}\begingroup\raggedright\endgroup

\end{document}